# Classification of cyber attacks on IoT and ubiquitous computing devices

Monika Freunek*, Alexandra Rombos*

*Abstract*—As the Internet of Things (IoT) has become truly ubiquitous, so has the surrounding threat landscape. However, while the security of classical computing systems has significantly matured in the last decades, IoT cybersecurity is still typically low or fully neglected. This paper provides a classification of IoT malware. Major targets and used exploits for attacks are identified and referred to the specific malware. The lack of standard definitions of IoT devices and, therefore, security goals has been identified during this research as a profound barrier in advancing IoT cybersecurity. Furthermore, standardized reporting of IoT malware by trustworthy sources is required in the field. The majority of current IoT attacks continue to be of comparably low effort and level of sophistication and could be mitigated by existing technical measures.

## I. INTRODUCTION

THE first IoT device eased the life of the computer scientists at Carnegie Mellon University in the early eighties of the last century. For many researchers, the coke machine was several floors away, with the risk of walking to the machine only to find no or, even worse, lukewarm coke. Students and staff deployed sensors in the machine and connected it to a server that could be accessed via ARPANET, the predecessor of the internet [1]. A few years later, computer scientist Mark Weiser and his team at Palo Alto Research Center envisioned ubiquitous computing and a so-called Smart Matter, where computing systems of all sizes would be distributed everywhere and operate in a mostly autarkic manner [2]. As these systems would be barely noticed or even forgotten, Weiser called it the era of "silent computing", where distributed small computers would be ubiquitous and silently assist in a broad variety of functions. With the progress of the internet, the term Internet of Things (IoT) was coined. Other descriptions include pervasive, or edge computing.

In 2032, the estimated number of connected IoT devices ranged around 15 billion [3, 4]. At the same time, the number of attacks on IoT devices reached record levels and increased in scope and their range of targets [5, 6].

[1] "This work was supported by the Rogers Cybersecure Catalyst Fellowship program. *Both authors contributed equally. *(Corresponding author: Monika Freunek)*

Monika Freunek is a research fellow alumni of the Rogers Cybersecure Catalyst at Toronto Metropolitan University, Toronto, ON M5B 2K3 Canada, and the founder of Lighthouse Science Consulting and Technologies Inc., Nova Scotia, Canada (e-mail: monika.freunek@torontomu.ca; monika.freunek@lighthouse-sct.com ).

Alexandra V. Rombos is an IT and Commerce undergraduate student at Toronto Metropolitan University, Toronto, ON M5B 2K3 Canada. (e-mail: alexandra.rombos@torontomu.ca).

Up to today, a significant number of attacks on IoT devices benefit from low or even neglected product security. Common examples for this include the use of default passwords and missing encryption. A frequent attack vector uses the typically low network visibility of IoT systems as an entry into a connected network, such as in the hack of the Jet Propulsion Laboratory via a Raspberry Pi, that had not been registered and monitored in the network [7]. Many attacks do not target the IoT devices themselves. Instead, the low level of IoT security provides for low-effort options to achieve cyber attack goals. On the other hand, attacks on critical infrastructure, aiming for example at Supervisory and Control Data Acquisition (SCADA) and Industrial Control Systems (ICS) such as SCADA devices in energy utilities, include highly targeted specific tools [8]. The trend in targeting critical infrastructure is increasing [6], and so are the risks for societies as a whole.

Denial of service attacks, manipulation of data and ransomware attacks on a vast number of IoT devices can have potentially devastating effects on individuals but could readily be national in scale, be it economically or physically. Examples of these scenarios include attacks on heating systems or communication infrastructure devices, such as routers. These attacks and their resulting harm can be limited to individuals and single businesses, but can also extend to large-scale attacks with a significant extension of the overall impact on other systems.

Therefore, it is essential to develop a profound understanding of current IoT cybersecurity threats. From a research perspective, the field of IoT cybersecurity is still evolving. Especially in the last years, several studies have been published on the topic, however these focused mostly on selected specific questions [see, for example 9-15]. This article investigates the history and the state of the art of IoT cybersecurity in general and identifies the major issues and recommendations. It establishes a classification of existing IoT malware and identifies commonly exploited vulnerabilities and the main types of targeted IoT devices. The structure of this paper is as follows. First, in Section II, general definitions and the definition of an IoT device applied within this work are discussed. Section III outlines the history and development of IoT malware, including a "family tree". The most common vulnerabilities and exploits are evaluated in Section IV. Section V classifies groups of mainly targeted devices. The paper ends with a summary of the major findings of this work and recommendations for an effective improvement of IoT security at the current stage.

## II. DEFINITION OF IoT DEVICES

A classification of IoT malware and attacks requires a common definition of an IoT system. As the term IoT has evolved historically, there are various definitions of IoT systems. For example, a study by *Sorry et al.* identified 122 definitions [16].

The current lack of a common definition of IoT is a fundamental challenge for IoT cybersecurity, as it affects the ongoing tracking and classification of IoT malware and vulnerabilities and the effective scope of IoT standards. As the following sections will show, the variety of definitions of IoT results in a variety of definitions of what is counted as an IoT malware. Where cybersecurity measures and standards exist, such as in larger organizations, the distinction between classical IT and IoT often leads to IoT and OT systems remaining unprotected, or some IoT systems such as printers being treated with standard IT security measures. Due to the characteristics of IoT devices, such as low computing resources, installation locations that are open to physical access and manipulation and often little or no means for encryption, authentication and access management, classical IT security measures are only of limited applicability.

This section therefore discusses common interpretations of IoT devices to then clarify the definition of an IoT system used within this paper.

Since Weiser's vision of the era of silent computing and the coining of the term Internet of Things, definitions of IoT devices in scientific literature and regulations have ranged from very limited to very broad interpretations. Components of computers and IoT devices, such as data storage, real-time clocks or analog-digital converters, are outside of the major scope of this work and will be neglected where applicable. Three definitions will be presented and discussed in the following.

**IoT Definition 1** *Systems, that include a microprocessor or microcontroller, a means of communication and a sensor or actor.*

**IoT Definition 2 a** *Any computing system that is not a so-called standard computer.* This definition obviously requires a definition of a standard computing device. Usually, these include desktop computers, laptops and servers. Laptops, however, can also be used as IoT devices, by means of their integrated or specifically connected sensors. Similarly, integrated laptop cameras can be operated as security cameras by using IoT protocols. Desktop computers may have external cameras connected, which are IoT devices. Finally, devices such as wireless headsets, earphones, computer mice or keyboards are often perceived by their users as an integrated part of their standard computing working stations. Technically, however, they are IoT devices which effectively means standard computer systems are at least partially IoT systems. This yields the more general description of this
**definition 2 b** *Devices that are connected with a communication technology to other computing devices, are addressable, and can process data without live human intervention.*

**Definition 3** In industry and technical literature, the IoT is frequently described as *devices that are connected to the internet and that perform tasks in the automation of industrial and agricultural processes (Industry 4.0) or within smart home environments*.

Figures 1-3 show schematics of the aforementioned three system interpretations.

Many common interpretations of the term IoT do not include devices from Industrial Control Systems (ICS) or operational technologies (OT). In general, these were originally not designed to be connected with the internet or other computing systems, but for isolated applications and environments, such as SCADA systems in energy utilities. These devices fall within the general definitions of (1) and (2), but not definition (3). The same is valid for devices such as routers, printers or smartphones. Because smartphones can also be used for standard computing tasks with direct human interaction such as office applications, they are regularly not classified as IoT devices. Smartphones, however, easily carry more than 10 sensors for a variety of classical IoT applications, such as location tracking, velocity measuring, light detection, imaging and sound processing. We therefore argue that smartphones should be classified both as standard and as IoT systems.

On the other hand, the IoT devices of the more narrow definitions 1 and 3 already vary significantly in their final design and in their communication technologies, computing power, energy demand, and physical exposure. All of this affects the range of potential vulnerabilities and the respective mitigation measures.

For these reasons, within this work all devices that are not a standard computer are considered as IoT devices. No distinction is made between IoT and OT despite where it is necessary by the context. This understanding aligns with definition 2.

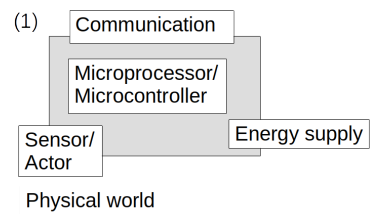

**Fig. 1.** Schematic of an IoT device, interpretation (1): generic system understanding where an IoT device includes a microprocessor or microcontroller, a means of communication and a sensor or actor.

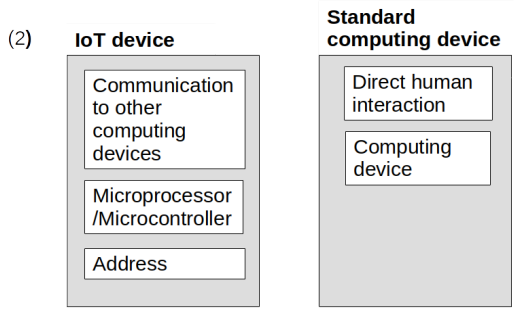

**Fig. 2.** Schematic of the IoT device interpretation (2): All computing systems that are not standard computers such as desktop computers or servers are considered as IoT devices.

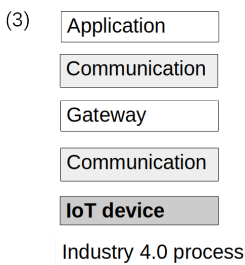

Industry 4.0 process

**Fig. 3.** Schematic of an IoT device, interpretation (3): devices that interact with the environment and applications in smart environments, such as smart homes.

## III. HISTORY OF IOT MALWARE

The following section provides an overview of the most common and a selection of especially notable IoT malware.

Other than in classical IT security, many IoT threats are not classified by a standardized and centralized approach that is maintained by at least one public source of a standing credibility, such as government agencies. An example is the *Common Vulnerabilities and Exposure (CVE)* inventories initiated by the *MITRE Corporation* under the US Cybersecurity and Infrastructure Security Agency (CISA) or specific alerts by governmental sources such as *NIST* and the *FBI* in the US, the *Canadian Center for Cybersecurity* or the annual *Threat Landscape Report* by the *European Union Agency for Cybersecurity ENISA*. Together with the heterogeneous definitions of the IoT, the missing coordinated and structured reporting of IoT vulnerabilities poses an additional challenge in establishing and maintaining a taxonomy of IoT malware. In this work we have used a variety of publicly available sources such as from academia, industry and governmental agencies. The majority of available sources were reports by cyber security companies as well as cyber security magazines. The most elaborate source was the *New Jersey Cybersecurity & Communications Integration Cell* by the State of New Jersey, which contains a detailed threat profile of major IoT malwares and attacks. This source was especially helpful in discovering the history of a malware and its basic features.

One of the first key IoT malware pieces that set the foundations for wide-scaled attacks and further developments until today is *Hydra* [17, 18]. Published as open source in 2001, the program provided means to parallelize brute force cracking of passwords. As weak or missing passwords are still one of the major issues in IoT security, this tool became one of the foundations of IoT malware. Several successors and further developments followed. In the context of IoT, *Hydra-2008.1* is the most notable, as it is considered to be the first malware specifically targeted at IoT devices, notably *D-Link* routers [19].

As can be seen on Figure 4, *Linux.Hydra* had a number of descendants and its influence can be seen throughout the family tree of IoT malware. *Psyb0t* and *Chuck Norris* came soon after and jointly influenced *Tsunami,* also known as *Kaiten,* in 2010.

Today's populated and complex IoT family tree stems mainly from three key botnet "ancestors": *Hydra*, *Mirai*, and *Bashlite*. Each ancestor was labeled as the root of the tree for its early appearance in the IoT attack scene and relates to the descendants that have stemmed from it.

While it is common for malware to use vulnerabilities and their exploits, the methodology used with each variant slightly differs between the families and branches. It is also important to note that while many variants descended from only one ancestor, some ancestors were combined to create different variants. One example of this is *EnemyBot* which has traces of both *Mirai* and *Bashlite* in its source code [20].

The variants at the bottom without a branch attached do not have an identified primary ancestor, but have still played a significant role in the history of IoT malware and attacks. For instance, *Leet* was discovered only months after *Mirai*, but has been claimed to have some key differences that do not make it a descendant of *Mirai*. *Leet* differs from *Mirai* in terms of payload, as *Mirai* was not originally coded to carry out huge SYN attacks (especially of those reaching 650 Gbps) and has its payloads hard coded differently than *Leet* [21]. Another key malware to note is *Stuxnet*. This IoT worm emerged in 2010 and used zero-day exploits and specifically targeted ICS code to sabotage nuclear facilities in Iran [22]. For some, Stuxnet marked the beginning of cyberwar aimed at critical infrastructure [23]. It is also likely the first IoT malware to get past an air-gapped system using USB sticks to initially infect the facility.

*Bashlite* is the ancestor highlighted in yellow at the top of the family tree and goes by a lengthy list of pseudonyms including "*Gafgyt*", "*QBot*", and "*LizardStresser*" [24]. It received its name from its use of the *Bash Bug*, or *Shellshock* vulnerability [25]. The *Bashlite* malware looks for routers connected to an open network [26] and can gain unauthorized access through logging into *BusyBox,* a unix utility package [27]. It was found to have infected about one million devices worldwide as of September 2016 [28] and has undoubtedly grown exponentially since the source code was fully published in 2017 [29].

Similar to the other key ancestors in the IoT Malware family tree, *Bashlite* has many variants both of its own and with other ancestors. Most notable is *EnemyBot* which was created combined with *Mirai* as well as *KTN-Remastered* stemming from *Kaiten* and *Bashlite*.

As seen bolded in red, *Mirai* is a botnet that infects devices such as home routers, network-enabled cameras and digital video recorders by going through a list of common default usernames and passwords for these devices [30, 31]. *Mirai* was first discovered in 2016 and it was used to launch a DDOS attack against the Krebs on Security site, a popular blog that educates readers on cybersecurity news and best practices [30, 32]. Despite users changing their device's credentials, they could still be at risk of getting attacked by *Mirai*. "...even if one changes the password on the device's Web interface, the same default credentials may still allow remote users to log in to the device using telnet and/or SSH." [33].

The 2016 *Mirai* attack revolutionized the way hackers take advantage of unsecured IoT devices as once the source code was leaked, it paved the way for future cybercriminals to collaborate and build on each other's work. As indicated by [34], the release of its source code "... gave birth to a wide variety of new malwares based on *Mirai*, often more sophisticated and with improved capabilities". Only a few significant *Mirai* variants are shown and used as a sample in the family tree, but in today's digitized era the 2016 botnet has created many known variants and likely countless undiscovered ones.

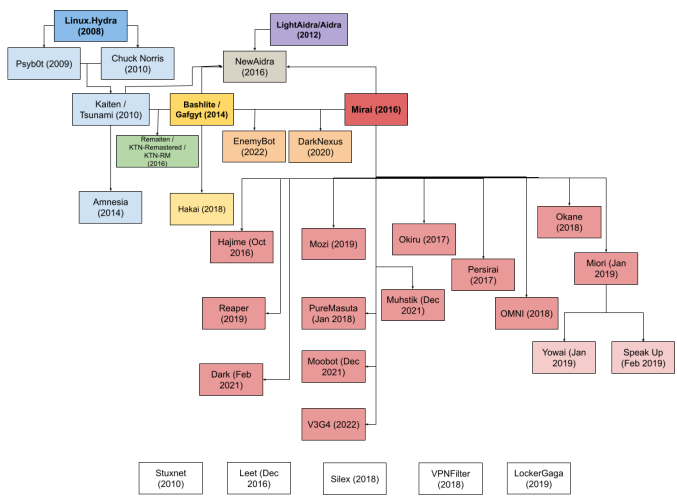

**Fig. 4.** Family tree of IoT malware. The year numbers indicate the first identified year of reporting of the malware.

## III. COMMONLY USED EXPLOITS

The low level of cybersecurity, the widespread use and the heterogeneity of IoT devices pave the way for a variety of devices to be susceptible to attacks. Any kind of network-connected piece of machinery is vulnerable and likely has been attacked before.

Table I summarizes the most frequently used major exploits in IoT malware. Where applicable, CVE numbers are provided. Two key findings can be derived from this evaluation. First, the kind of vulnerabilities that are being used to access and attack IoT devices are not overly sophisticated. There are no complex mechanisms or ploys to sneak into a device, but rather using a combination of guessing a default or weak password and then typically taking advantage of a remote code execution exploit. If successful, the device is then at the mercy of the malware and it can be used to do anything the attacker pleases. From being a part of an army botnet, launching DDOS attacks on websites, or accessing private and critical information, an IoT device can give a threat actor a plethora of resources and tools to cause harm. The vulnerabilities have been known in some cases for more than a decade, yet they still provide for the majority of IoT attacks. On top of that, almost 10 years after its discovery, Mirai still accounts for most IoT attacks, while the rest stem from other mostly known IoT malware [42, 43].

TABLE I
Most frequently used major exploits and CVE numbers in common IoT malware.[2]

| Malware | Exploit applied | CVE number |
| --- | --- | --- |
| Mirai [30] | default login credentials | N/A |
| Bashlite [26], [35], [36] | code injection, remote code execution | CVE-2014-6271 CVE-2014-8361 |
| Hajime [37], [38] | TR-069 (open port on a router), default password, Arris cable modem password of the day | N/A |
| VPNFilter [39] | JavaScript injection | N/A |
| Reaper [40] | default password code injection | N/A |
| Persirai [36], [41] | default password code injection | CVE-2014-8361 |
| Kaiten/ Tsunami [18] | remote code execution | N/A |

[2] *CVEs are often device or software specific. Incomplete or competing information was omitted.*

## III. CLASSIFICATION OF ATTACKED DEVICES

A similar situation presents itself when investigating common types of attacked devices as shown in Table II. Again, given the age of the majority of the malware, the result remains remarkably unchanged with routers, cameras, and DVRs being the main targets. While this can be partially caused by the code of some of the malware, the result shows that IoT cyber risk mitigation measures remain low or neglected and have not been adapted to the known threats and the increased threat landscape.

TABLE II
Main devices attacked by IoT malware.

| Malware | Device |
|---|---|
| Mirai | home routers, CCTV cameras, DVRs |
| Bashlite | CCTV cameras, DVRs, home routers |
| Hajime | DVRs, cameras, routers |
| VPNFilter | routers, NAS devices |
| Reaper | CCTV cameras, DVRs, routers |
| Persirai | IP cameras |
| Kaiten/Tsunami | routers, cameras, DVRs |
| LightAidra | routers, TV, DVR, VOIP devices, cameras |

Fig. 6. Common devices attacked by IoT malware

The most common types of IoT devices attacked are routers and cameras. These devices are especially popular in homes and businesses, which makes them the perfect target in places where cybersecurity is not a major concern for users.

Routers and cameras may also contain personally identifiable data which should be a top priority to protect. Not only is this data susceptible to being exposed, but it could also put the user at risk of identity theft, losing access to their financial details, and having all personal information stolen. The most concerning aspect of the IoT botnet attacks is when a device is compromised, it usually isn't obvious to the user as it is carrying out malicious acts in the background while the IoT device appears to run as usual. This means that emphasis should be placed on routinely scanning IoT devices and re-booting them, as well as changing the login credentials whenever there is a suspected compromise.

## V. OUTLOOK AND CONCLUSIONS

As shown through the classification of attacks and the family tree, IoT attacks are getting more expansive which further proves the need to secure all aspects of IoT devices.

Yet, as shown in Table I and Table II, at the current stage relatively simple measures such as regularly rebooting routers, using proper passwords, and blocking unused ports significantly mitigate the risks associated with using an IoT device. The technical analysis of the IoT threats and the required protection measures are widely known, yet commonly not applied, even in commercial products such as routers. This brings up the role of non-technical factors in IoT cybersecurity such as development and adherence to standards, accountability within organizations and manufacturers, as well as controls and regulations established by governments.

Another challenge that was identified during this research is the lack of basic IoT cybersecurity research and standards. As discussed in the first sections of this paper, this starts with the various definitions of IoT devices, which in turn provides for varying and incomplete results and outcomes from technical and non-technical IoT cybersecurity research and measures.

In addition to this, reliable information on the origin of IoT malware and credible, standardized, and current data repositories on IoT malware are lacking and required for efficient IoT cybersecurity research and measures, especially those relying on automation or machine learning.

At the present stage, obtaining a fundamental data base on IoT malware requires own and comparably extensive research and validation, as existing sources are regularly contradicting and their credibility is not always verifiable. This in turn makes further validation and often own research necessary. The majority of available resources do not meet scientific standards. However, for lack of alternatives they are commonly used even in research work as the paper at hand.

There is a need for a standardized verification process that ensures the information being provided is accurate. For example, the initial findings of the malware are typically published on independent research blogs and the source code for many variants is on Github - both of which have a track record of being illegitimate sources until the information is verified by a larger and more reputable company. Much historical and technical information about IoT malware is vastly spread out across journals and websites on the internet, and requires many hours of searching to obtain enough clear information on one variant. Even when information on the malware is found, it is often reported using different names, which are also not always referenced to each other. A typical example for this is *Bashlite.* As a result, researchers may oversee a specific pseudonym, or mistake the same malware as two different variants instead of one.

In order to provide IoT cybersecurity research with a solid foundation to advance the field profoundly, a current and standardized data repository of IoT malware is required from an official credible organization, such as a government institution. The New Jersey Cybersecurity and Communications Integration Cell with its search function to find any recorded details of specific variants of IoT malware, is a good example, yet it still partially misses the origination and technical details of the reported malware [46].

The current lack of IoT cybersecurity research seems to run in contradiction to its importance for societies of today and reflects the state of applied IoT cybersecurity. Without a significant increase in research on the foundations of IoT cybersecurity as outlined in this work, the success of urgently required nontechnical measures such as IoT standards and regulation will be limited.

ACKNOWLEDGMENT

The authors would like to thank all members of the Rogers Cybersecure Catalyst at Toronto Metropolitan University, especially Prof. Marcus Santos and Dr. Randy Purse for their ongoing support during this work and their review of this paper.

REFERENCES

[1] www.cs.cmu.edu/~coke, assessed 28-02-2023.
[2] Weiser, M. (1991). The computer for the 21st Century. Scientific American, 265(3), 75-84.
[3] https://www.statista.com/statistics/1183457/iot-connected-devices-worldwide/, accessed 19-10-2023.
[4] https://iot-analytics.com/number-connected-iot-devices/, accessed 19-10-2023.
[5] https://www.sonicwall.com/medialibrary/en/white-paper/mid-year-2023-cyber-threat-report.pdf, accessed 19-10-2023.
[6] https://hub.dragos.com/hubfs/312-Year-in-Review/2022/Dragos_Year-In-Review-Report-2022.pdf?hsLang=en, accessed 19-10-2023.
[7] https://www.cigionline.org/articles/the-ungoverned-space-of-us-space-cyber-governance/, accessed 19-10-2023.
[8] https://www.cisa.gov/news-events/cybersecurity-advisories/aa22-103a accessed 19-10-2023.
[9] Mogadem, M.M., Li, Y. & Meheretie, D.L. A survey on internet of energy security: related fields, challenges, threats and emerging technologies. *Cluster Comput* 25, 2449–2485 (2022)..
[10] V. Hassija, V. Chamola, V. Saxena, D. Jain, P. Goyal and B. Sikdar, "A Survey on IoT Security: Application Areas, Security Threats, and Solution Architectures," in *IEEE Access*, vol. 7, pp. 82721-82743, 2019
[11] N. Mishra and S. Pandya, "Internet of Things Applications, Security Challenges, Attacks, Intrusion Detection, and Future Visions: A Systematic Review," in *IEEE Access*, vol. 9, pp. 59353-59377, 2021.
[12] Williams, P., Dutta, I.K., Daoud, H. & Bayoumi, M. 2022, "A survey on security in internet of things with a focus on the impact of emerging technologies", *Internet of Things,* vol. 19, pp. 100564.
[13] Hossain, M.M., Fotouhi, M. and Hasan, R., 2015, June. Towards an analysis of security issues, challenges, and open problems in the internet of things. In *2015 ieee world congress on services* (pp. 21-28). IEEE.
[14] Schiller, E., Aidoo, A., Fuhrer, J., Stahl, J., Ziörjen, M. & Stiller, B. 2022, "Landscape of IoT security", *Computer science review,* vol. 44, pp. 100467.
[15] Cirne, A., Sousa, P.R., Resende, J.S. & Antunes, L. 2022, "IoT security certifications: Challenges and potential approaches", *Computers & security,* vol. 116, pp. 102669.
[16] K. Sorri, N. Mustafee, and M. Seppänen, "Revisiting IoT definitions: A framework towards comprehensive use," *Technological forecasting & social change*, vol. 179, p. 121623, 2022.
[17] THC, Hydra, https://github.com/vanhauser-thc/thc-hydra, accessed 19-10-2023.
[18] M. De Donno, N. Dragoni, A. Giaretta, A. Spognardi, "DDoS-Capable IoT Malwares: Comparative Analysis and Mirai Investigation", *Security and Communication Networks*, vol. 2018, Article ID 7178164, 30 pages, 2018.
[19] A. Costin, J. Zaddach, "IoT malware: comprehensive survey, analysis framework and case studies", Proc. Black Hat USA, 2018.
[20] Paganini, P. (2022) EnemyBot malware adds new exploits to target CMS servers and Android devices, Security Affairs. Available at: https://securityaffairs.co/131783/malware/enemybot-botnet-new-exploits.html, accessed 19-10-2023.
[21] Bekerman, D. and Zawoznik, A. (2016) 650 Gbps ddos attack from the Leet botnet, Imperva. Available at: https://www.imperva.com/blog/650gbps-ddos-attack-leet-botnet/?redirect=Incapsula, accessed 19-10-2023..
[22] D. Kushner, "The real story of stuxnet," *IEEE spectrum*, 50(3),. pp. 48–53, 2013.
[23] T. M. Chen, "Stuxnet, the real start of cyber warfare? [Editor's Note]," *IEEE network*, 24(6), pp. 2–3, 2010.
[24] Cyber Alerts: BASHLITE Botnet (2020) NHS Digital. NHS. Available at: https://digital.nhs.uk/cyber-alerts/2018/cc-2557, accessed 19-10-2023.
[25] Anonymous, "Taiwan Province of China : Trend Micro Launches Free Protection for Shellshock a.k.a. Bash Bug," *MENA Report*. September 2014.
[26] Schick, S. (2014) Latest shellshock attack uses bashlite to target devices running BusyBox, Security Intelligence. IBM. Available at: https://securityintelligence.com/news/latest-shellshock-attack-uses-bashlite-target-devices-running-busybox/, accessed 19-10-2023.
[27] Schick, S. (2019) Wireless routers exploited by Gafgyt variant could be used in ddos attacks, Security Intelligence. IBM. Available at: https://securityintelligence.com/news/wireless-routers-exploited-by-gafgyt-variant-could-be-used-in-ddos-attacks/, accessed 19-10-2023.
[28] Loeb, L. (2016) Bashlite malware uses IOT for ddos attacks, Security Intelligence. IBM. Available at: https://securityintelligence.com/news/bashlite-malware-uses-iot-for-ddos-attacks/ , accessed 19-10-2023.
[29] NJCCIC (2016) NJCCIC Threat Profile - Bashlite, Cyber.nj.gov. New Jersey Cybersecurity & Communications Integration Cell. Available at: https://www.cyber.nj.gov/threat-center/threat-profiles/botnet-variants/bashlite (Accessed: April 5, 2023).
[30] Malware Must Die Blog, "MMD-0056-2016 - Linux/Mirai, how an old ELF malcode is recycled..", September 2016, https://blog.malwaremustdie.org/2016/08/mmd-0056-2016-linuxmirai-just.html, accessed 19-10-2023.
[31] E. Bertino and N. Islam, "Botnets and Internet of Things Security," in Computer, 50(2), 76-79, Feb. 2017.
[32]Krebs, B. (2016) Krebsonsecurity hit with record ddos, Krebs on Security. Available at: https://krebsonsecurity.com/2016/09/krebsonsecurity-hit-with-record-ddos/ , accessed 19-10-2023.
[33] Krebs, B. (2016) Who makes the IOT things under attack?, Krebs on Security. Available at: https://krebsonsecurity.com/2016/10/who-makes-the-iot-things-under-attack/ accessed 19-10-2023.
[34] De Donno, M. et al., "DDoS-capable IOT malwares: Comparative Analysis and Mirai Investigation", Security and Communication Networks, 2018.
[35] CVE (2014) CVE-2014-6271, CVE. The MITRE Corporation. Available at: https://cve.mitre.org/cgi-bin/cvename.cgi?name=CVE-2014-6271, accessed 19-10-2023.
[36] CVE (2014) CVE-2014-8361, CVE. The MITRE Corporation. Available at: https://cve.mitre.org/cgi-bin/cvename.cgi?name=CVE-2014-8361, accessed 19-10-2023.
[37] S. Edwards, I. Profetis, "Hajime: Analysis of a decentralized internet worm for IoT devices", Rapidity Networks, 16, pp. 1-18, 2016.
[38] NJCCIC (2016) NJCCIC Threat Profile - Hajime Botnet, Cyber.nj.gov. New Jersey Cybersecurity & Communications Integration Cell. Available at: https://www.cyber.nj.gov/threat-center/threat-profiles/botnet-variants/hajime-botnet, accessed 19-10-2023.
[39] J. C. Sapalo Sicato, P. K. Sharma, V. Loia, and J. H. Park, "VPNFilter Malware Analysis on Cyber Threat in Smart Home Network," *Applied sciences*, 9(13), p. 2763, 2019.
[40] Anonymous "Govt issues advisory on 'Reaper' botnet," *The Hitavada*, 2017.
[41] A. Pandey, "Trend Micro Detects Persirai IoT Botnet That Targeted 120,000 IP Cameras," *PC quest : the personal computing magazine*, 2017.

[42] H. Sato, H. Inamura, S. Ishida, and Y. Nakamura, "Plain Source Code Obfuscation as an Effective Attack Method on IoT Malware Image Classification," IEEE, 2023, pp. 940–945.

[43] M. Dib, S. Torabi, E. Bou-Harb, and C. Assi, "A Multi-Dimensional Deep Learning Framework for IoT Malware Classification and Family Attribution," *IEEE eTransactions on network and service management*, vol. 18, no. 2, pp. 1165–1177, 2021.

[44] NJCCIC (2015-2023) NJCCIC, Cyber.nj.gov. New Jersey Cybersecurity & Communications Integration Cell. Available at: https://www.cyber.nj.gov, accessed 19-10-2023.

**Dr.-Ing. Monika Freunek** is a Research Fellowship Alumni of the *Rogers Cybersecure Catalyst Fellowship Program* with *Toronto Metropolitan University,* and the founder of *Lighthouse Science Consulting and Technologies Inc.*, Canada. She holds a PhD in microsystems engineering from the *Albert-Ludwigs-University* of Freiburg, Germany, and is an experienced researcher and lecturer, project manager and executive leader in the fields of critical infrastructures and cybersecurity. She has published more than 50 articles, book chapters, books and patents in the fields of energy and IoT systems, microsystem technologies, cybersecurity, data science and machine learning. Her fellowship project partially presented in this work focused on available approaches to IoT cybersecurity in research and industry and the theoretical limits of achievable cybersecurity in distributed systems.

**Alexandra Rombos** is a 5th year IT and Commerce undergraduate student at Toronto Metropolitan University, in Toronto, and participated as a research assistant in the Rogers Cybersecure Catalyst Research Fellowship Program. She is passionate about technology and is pursuing a minors in Computer Science as well. In the field of IoT cybersecurity, Alexandra seeks to understand the history and present-day landscape of IoT technologies and how people can best equip themselves to avoid being targeted in the future.